\documentclass[sigconf]{acmart}

\renewcommand\footnotetextcopyrightpermission[1]{} % removes footnote with conference information in first column
\setcopyright{none}
\acmConference[ISE 2026]{a}{September 11, 2026}{São Paulo, SP, Brazil}

\AtBeginDocument{
    
}

\usepackage{booktabs}   % professional tables
\usepackage{balance}    % balance columns on the last page

\begin{document}

%% allowing the author to define a "short title" to be used on the page 
%% headers.
\title{Building AI-Intensive Software \emph{with} AI:\\
Early Results and a Cautionary Tale on Measuring Development Cost}

\author{Victor Barros de Miranda Neves}
\affiliation{
  \institution{Centro de Informática, Universidade Federal de Pernambuco}
  \city{Recife}
  \country{Brasil}
}
\email{vbmn@cin.ufpe.br}

\author{Kiev Santos da Gama}
\affiliation{
  \institution{Centro de Informática, Universidade Federal de Pernambuco}
  \city{Recife}
  \country{Brasil}
}
\email{kiev@cin.ufpe.br}

\author{Vinicius Cardoso Garcia}
\affiliation{
  \institution{Centro de Informática, Universidade Federal de Pernambuco}
  \city{Recife}
  \country{Brasil}
}
\email{vcg@cin.ufpe.br}

%% By default, the full list of authors will be used on the page
%% headers. This list is often too long and will overlap
%% other information printed in the page headers. 
%% This command allows the author to define a more concise list
\renewcommand{\shortauthors}{Neves et al.}
\renewcommand{\shorttitle}{Building AI-Intensive Software \emph{with} AI}

%% The abstract is a short summary of the work the paper presents.
\begin{abstract}
Empirical reports on the true cost of AI-intensive software development remain scarce, and the few that exist are easy to get wrong in ways that never surface in the final number. We report early results from an ongoing case study: a six-person student team built a full conversational onboarding assistant — RAG-based code chat, guided tours, dependency graphs, technical-debt analysis — over one academic term using pervasive AI assistance. We instrumented development with a three-layer cost model (real AI spend, self-reported human effort, human counterfactual) and initially reported a 19.4$\times$ cost ratio. A follow-up pass revealed two independent errors — inferring per-token cost under a flat-rate subscription, and pricing the counterfactual with the wrong regional labor rates — that together had inflated the ratio by roughly 2$\times$; the corrected figure is $\sim$9.9$\times$. We present this correction as an early, generalizable finding in its own right: both errors are easy to make, invisible in the final number, and plausibly common in similar reports. We outline next steps toward a more robust, replicable costing methodology for AI-intensive development.
\end{abstract}

\keywords{AI-Assisted Software Development, Generative AI, Software Economics, Retrieval-Augmented Generation, Developer Onboarding}
 
\maketitle
 
%% ============================================================
%% 1. INTRODUCTION
%% ============================================================
\section{Introduction}
\label{sec:intro}
 
Large language models now execute complex software engineering tasks end-to-end --- writing code, generating tests, drafting documentation --- but most evidence evaluates isolated tasks rather than complete systems built under pervasive AI assistance \cite{fan2023llm4se, hou2024llm4se}. Coding assistants have moved from single-line completion to agentic modes that plan, edit multiple files, and iterate, but open questions remain around correctness, review burden, and the economics of adoption \cite{sauvola2024genai}. Broader reviews of AI for software engineering explicitly call for more evidence on real-world adoption and outcomes \cite{ahmed2025ai4se}, and this call is sharpest for cost: practitioners lack empirical baselines for how much effort can realistically be delegated to an AI agent, and the few cost figures that circulate \cite{paradis2025much, faruqui2024ai} are rarely audited for measurement error.

The setting is further complicated when the system under construction is itself AI-intensive. Building a tool that heavily relies on LLMs and retrieval pipelines creates a doubly recursive scenario: engineers use AI to build software whose core behavior also depends on AI. We chose to study this scenario through a concrete case: a conversational onboarding assistant that helps developers navigate legacy codebases via retrieval-augmented generation (RAG) \cite{nam2024llmcode}, addressing a well-documented technical bottleneck in developer onboarding \cite{Buchan2019, Gregory2022, Ju2021} and developer experience more broadly \cite{Fagerholm2012, Greiler2023}.

We report emerging results from this ongoing case study. A six-person student team built the assistant over one academic term, relying almost entirely on AI assistance for both design and implementation. We instrumented the project with a three-layer cost model --- real AI spend, self-reported human effort, and an estimated human counterfactual --- as a first step toward a general costing methodology for AI-intensive development.

The most important finding was not the cost ratio itself but how fragile it turned out.  The first analysis reported a 19.4$\times$ ratio, later corrected to 9.9$\times$ after we identified two independent, easy-to-miss errors. We present this correction, our  methodology, and early lessons as preliminary evidence for a broader claim we are continuing to investigate: that cost claims about AI-assisted development deserve more scrutiny than the literature currently gives them.

\section{Background and Related Work}
\label{sec:related}
 
\noindent\textbf{AI-assisted software development.} The use of LLMs to assist software engineering tasks has grown rapidly, and recent surveys map an expanding landscape of applications from requirements to maintenance~\cite{fan2023llm4se,hou2024llm4se}. Coding assistants operating at the task level, instead of merely providing single-line completions, introduced ``agent'' modes that plan, edit multiple files, run commands, and iterate. While reported benefits center on implementation throughput, open questions remain around correctness, review burden, and the economics of adoption~\cite{sauvola2024genai}. Because most of this literature evaluates model capability on benchmark tasks, our work provides a complementary perspective by analyzing the process and cost of building a complete system pervasively using AI assistance.
 
\noindent\textbf{Program comprehension and developer onboarding.}
Understanding an unfamiliar codebase is a long-standing bottleneck. Industrial tools target parts of this problem, but they typically focus on pointwise questions or require manual authoring of walkthrough content. Research combining program comprehension with LLMs demonstrates that models can explain code and answer repository questions when fed adequate context via RAG~\cite{nam2024llmcode}. Codebase health is also a recognized and actionable factor in developer experience~\cite{Greiler2023}. Onboarding literature often focuses on social integration in agile settings~\cite{Buchan2019,Gregory2022}, but our domain specifically addresses the technical axis of onboarding. The system described in this paper serves as the engineering case study to ground our emerging results.
 
\noindent\textbf{Cost Measurement in Practice.}
Broad reviews of AI for software engineering call for more evidence on real-world adoption and outcomes~\cite{ahmed2025ai4se}. Transparent, audited cost measurements of AI-assisted development remain uncommon, particularly for AI-intensive products built in resource-constrained academic settings. Existing evidence rarely audits its own measurement pipeline: productivity studies report time or throughput gains rather than dollar cost~\cite{paradis2025much}, and cost-focused studies report top-line percentages without exposing pricing or counterfactual assumptions~\cite{faruqui2024ai}. Our case study fills this gap by treating measurement error itself as a finding.
 
\begin{figure}[t]
  \centering
  \includegraphics[width=\linewidth]{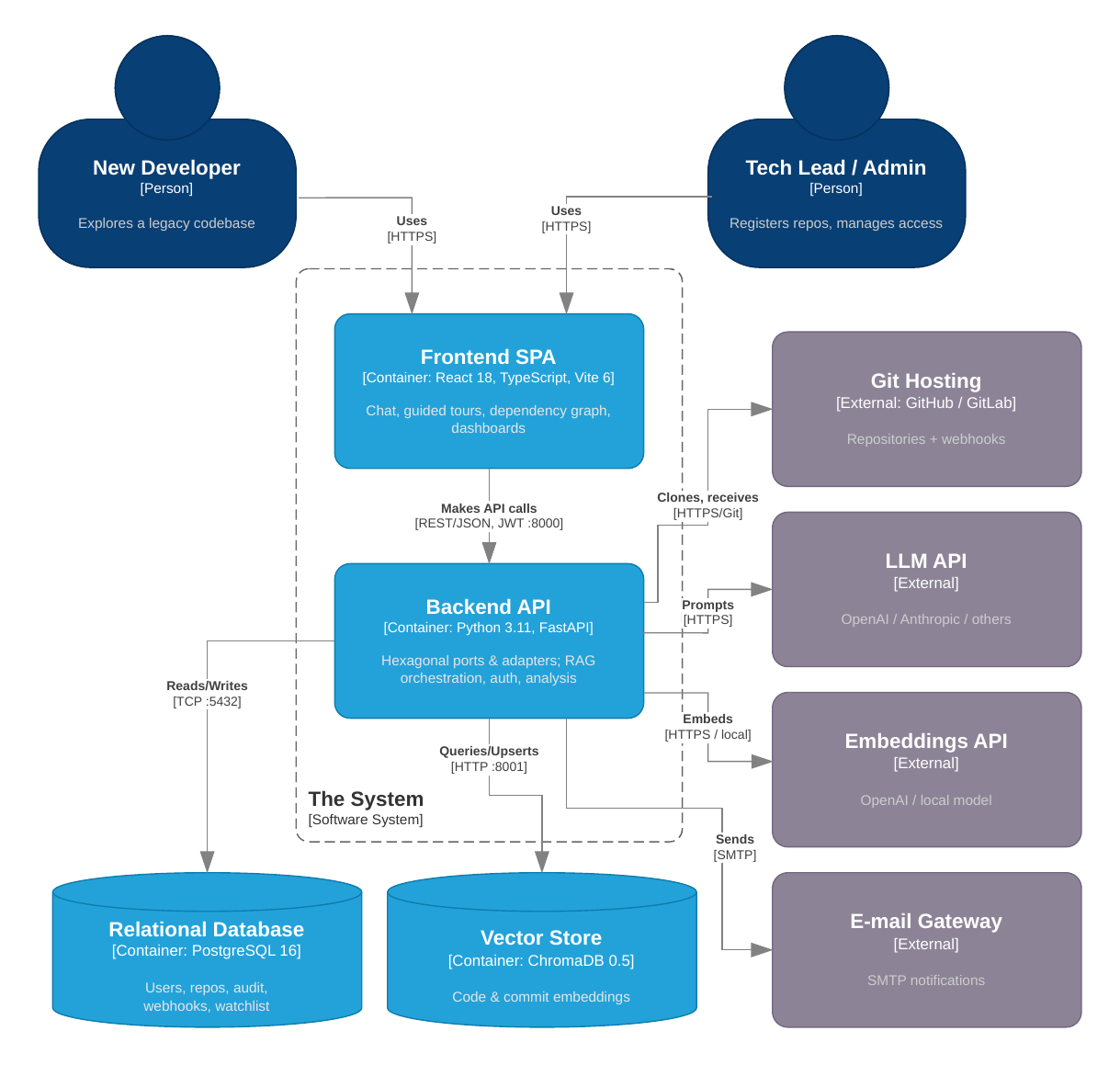}
  \caption{C4 container diagram: a React client calls a FastAPI backend (hexagonal ports and adapters) that orchestrates a vector store, relational DB, and external Git, LLM, embedding, and e-mail services. Edge labels give protocols and ports.}
  \label{fig:arch}
\end{figure}

\begin{figure}[t]
  \centering
  \includegraphics[width=\linewidth]{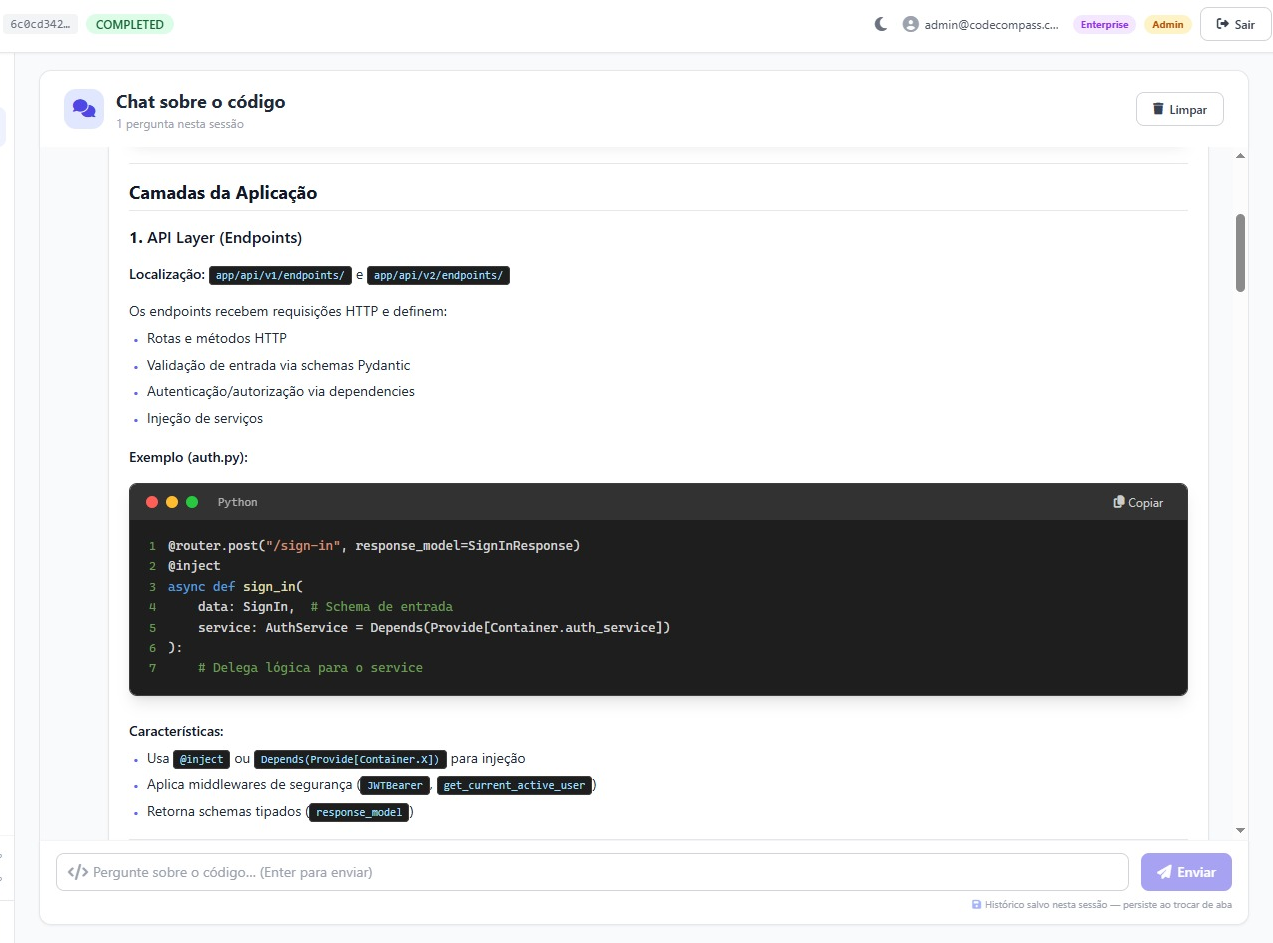}
  \caption{The RAG chat in use (UI in PT-BR). Answers are grounded in real repository fragments (here a router declaration, dependency injection, and a typed schema) rather than generic model knowledge.}
  \label{fig:ui}
\end{figure}
 
\begin{table}[t]
  \caption{Scale of the artifact. Line counts via \texttt{cloc} over
  hand-written source only (excluding dependencies, generated code, and
  configuration).}
  \label{tab:scale}
  \small
  \begin{tabular}{@{}ll@{}}
    \toprule
    \textbf{Dimension} & \textbf{Value} \\
    \midrule
    Features delivered & 25+ across 4 phases \\
    Supported languages & 15 (via tree-sitter) \\
    Automated tests & 201 cases in 26 files (unit, integration, e2e) \\
    Backend & Python, FastAPI, hexagonal architecture \\
    Frontend & React, Vite, TypeScript \\
    Retrieval & Vector store + sentence/OpenAI embeddings (RAG) \\
    LLM providers & Configurable (Anthropic / OpenAI / others) \\
    Lines of code & $\sim$21{,}000 production; $\sim$2{,}000 tests \\
    \bottomrule
  \end{tabular}
\end{table}

\section{Case and Method}
\label{sec:method}

\noindent\textbf{The system.} The team built a conversational onboarding assistant addressing a recognized gap where onboarding literature shows codebase orientation delivered largely through mentoring and ad hoc peer support~\cite{Ju2021,Fagerholm2012,Greiler2023}, drawing experienced developers away from their own work. The tool clones a Git repository, chunks source code with a multi-language parser (tree-sitter, 15 languages), embeds code and commit history into a vector store, and powers a RAG-based chat, automated guided tours, a module dependency graph, and technical-debt analysis grounded in real repository fragments rather than generic model knowledge. The backend follows a hexagonal architecture (Python/FastAPI); the frontend is a React single-page application. Over one academic term, the team delivered 25+ features and 201 automated tests ($\sim$21,000 production lines of code, $\sim$2,000 test lines).
 
\noindent\textbf{Development process.} Work proceeded through four phases (problem framing, solution design, build-and-test, and measurement) using two AI tools: a frontier-model LLM assistant for early-phase design research, and a mid-tier AI coding agent for implementation, testing, and documentation.
 
\noindent\textbf{Three-layer cost model.} For each phase, we recorded: (1) real AI spend, taken from provider billing records (a flat-rate coding-agent subscription plus metered charges for repository analysis and embeddings); (2) self-reported human effort, hours logged by team members; and (3) a human counterfactual, a bottom-up estimate of the hours a professional of a given seniority (junior/mid/senior) would need to complete each recorded task without AI assistance, priced at regional hourly rates (US\$3.86/6.67/10.88) derived from local gross salaries. Seniority was assigned by task complexity, not by which student performed the work. The effort and counterfactual figures are self-reported estimates subject to hindsight bias.

\section{Results and Findings}
\label{sec:results}
 
Comparing the project's actual cost with its estimated cost without AI reveals a large gap, but the key question is where that gap comes from, and where our initial estimation went wrong.
 
\noindent\textbf{Cost across phases.} Table~\ref{tab:econ} consolidates the three layers. Tooling cost US\$69 in total: a flat GitHub Copilot Business subscription (US\$19/month over three months, apportioned across phases), a metered repository-analysis service (US\$12.13), and embedding generation (US\$0.13). The team logged 35.2 hours of  human effort. Valuing that effort at the mid level rate gives US\$235, for a total project cost of \textit{US\$304}. The counterfactual (329 professional hours at the seniority profile each activity demands) comes to \textit{US\$3,005}. The resulting ratio is \textit{$\sim$9.9$\times$}.The distinction between the two effort figures matters and is easy to conflate: the 329 hours are an estimate of what a professional team would have needed \emph{without} AI; the team's own logged effort was 35.2 hours, roughly nine times less.
 
\begin{table*}[t]
\caption{Cost model per phase. Tool costs from billing records (flat-rate Copilot apportioned by usage); human effort self-reported at the mid-level rate (US\$6.67/h); counterfactuals are retrospective estimates. Columns figures were rounded.}
  \label{tab:econ}
  \small
  \begin{tabular}{@{}lrrrrrrr@{}}
    \toprule
    \textbf{Phase} & \textbf{Tools (US\$)} & \textbf{Human (h)} &
    \textbf{Human (US\$)} & \textbf{Total w/ AI (US\$)} &
    \textbf{Counterfact. (h)} & \textbf{Counterfact. (US\$)} &
    \textbf{Ratio} \\
    \midrule
    Pre-proposal            & 6   & 7.0  & 47    & 53    & 31  & 256     & 4.9$\times$ \\
    P1 (problem framing)    & 18  & 11.4 & 76    & 94    & 136 & 1{,}302 & 13.9$\times$ \\
    P2 (solution design)    & 11  & 4.2  & 28    & 39    & 19  & 186     & 4.8$\times$ \\
    P3 (build-and-test)     & 11  & 3.8  & 25    & 36    & 27  & 226     & 6.3$\times$ \\
    P4 (measurement)        & 12  & 5.8  & 39    & 51    & 67  & 598     & 11.8$\times$ \\
    Quality \& improvements & 12  & 3.0  & 20    & 32    & 49  & 436     & 13.6$\times$ \\
    \midrule
    \textbf{Total} & \textbf{69} & \textbf{35.2} & \textbf{235} &
    \textbf{304} & \textbf{329} & \textbf{3{,}005} & \textbf{9.9$\times$} \\
    \bottomrule
  \end{tabular}
\end{table*}

\noindent\textbf{Adjustments to the Cost Model.}
\label{sec:error}
Our first analysis of these data reported a ratio of 19.4$\times$ and identified the problem-framing phase as consuming 68\% of the AI budget, which we interpreted as evidence that model selection dominates the economics of AI-assisted development. Both claims were wrong, and the reasons are instructive enough that we report them rather than quietly correcting them. The error had two parts. \textit{First}, we estimated AI cost by counting tokens and multiplying by published per-token rates for the model we believed the coding agent was using. It was not: the agent ran under a \emph{flat-rate} subscription, so token volume had no effect on what we paid. The US\$22.74 we attributed to the problem-framing phase. It appeared to consume two-thirds of the budget and led us to conclude that model choice drove cost---was an artifact of that assumption: a plausible figure computed from a pricing model that did not apply. Once billing records replaced estimates, tooling cost $\sim$US\$69 in total, distributed in proportion to each phase's duration instead of which model it used. \textit{Second}, we priced the counterfactual using hourly rates from national salary surveys skewed toward the largest metropolitan markets, roughly 85\% above rates in the region where the work was performed. Correcting to regional rates lowered the counterfactual from ~US\$5,678 to ~US\$3,005. The two corrections move in the same direction---both had inflated the apparent advantage of AI---and together they take the ratio from 19.4$\times$ to 9.9$\times$. We think 9.9$\times$ is the more useful number precisely because every term in it is now traceable to a billing record or a stated rate.
 
\noindent\textbf{Distribution of AI-assisted work.} 
\label{sec:distribution}
The team's self-assessed reliance on AI varied sharply by activity, and the gradient is itself the finding. Implementation-level work was mostly delegated: code writing and test generation were about 95\% AI-assisted, and debugging around 80\%. Work that pairs mechanical output with human intent sat in the middle: documentation near 85\% and requirements analysis near 70\%. Judgment-heavy work stayed predominantly human---prompt design was only about 40\% AI-assisted and architectural decisions about 20\%. Thus, AI removed implementation friction while the decisions shaping the system remained with the team; the tool amplified the engineers instead of replacing them.
 
\noindent\textbf{Benefits and Limitations of AI Assistance.} 
Qualitatively, the AI assistance was key to rapidly exploring unfamiliar domains and translating established design decisions into consistent, tested code, whereas the assistance introduced friction in three conditions: (1) the models occasionally generated overly optimistic effort estimates; (2) long agent sessions silently lost context, forcing the re-exploration of previously read files; and (3) implementing features directly from prompts without a written design led to significantly more rework compared to features built from explicit specifications.

\section{Discussion}
\subsection{Lessons Learned}
\label{sec:lessons}

\noindent\textbf{Pricing model verification precedes cost optimisation.} 
Our costliest error was not a cost but a misunderstanding of how cost was incurred: we assumed per-token billing for a premium model, when the coding agent in fact ran under a flat monthly subscription, where token volume has no effect on spend. Optimising token usage under a flat-rate plan is wasted effort; a metered service, conversely, requires exactly that discipline. We suggest establishing for each tool whether it is flat-rate, metered, or hybrid before optimising, reading billing dashboards rather than inferring cost from token counts, and recording metered charges as they occur rather than reconstructing them retrospectively.
 
\noindent\textbf{Explicit context management in long agent sessions.} 
Silent context loss was the most common source of wasted effort in our case: earlier context was compacted away without warning, prompting the agent to re-explore files or contradict earlier decisions. Two habits appeared to mitigate this: a short running state note before each subtask, and recording architectural decisions in repository files rather than relying on conversational continuity.
 
\noindent\textbf{Specification-first task definition.} 
In our observation, features implemented directly from conversational requests required more iterations than those based on explicit specifications; without a written design, the agent filled gaps with plausible but often incorrect assumptions. This suggests AI assistance amplified existing clarity rather than substituting for it.
 
\noindent\textbf{Prompt cataloguing as an engineering practice.} 
Treating core application prompts as versioned artifacts with a fixed template appeared to aid reuse and review, giving them a level of scrutiny comparable to source code.
 
\noindent\textbf{Security practices established from initial development.} 
The pace of AI-assisted development makes security debt easy to accumulate. We adopted upfront rules (e.g., no hard-coded credentials in code or agent transcripts, constant-time comparison for webhooks, centralized audit logging) as a precaution against this risk.
 
\noindent\textbf{Interpretation of the reported cost ratio.} 
A ratio of 9.9$\times$ is not a controlled measurement of productivity: it divides a numerator now grounded in billing records and time logs by a denominator that remains a retrospective estimate of work never performed. It is also a conservative figure, since the 329 counterfactual hours assume a senior developer already fluent in the stack; a mid-level profile would plausibly raise the ratio to 13--16$\times$. We report the lower estimate deliberately and treat the multiplier as an order-of-magnitude signal from a single case, not a benchmark.

\subsection{Practical Implications}
\noindent\textbf{For SE Practice.}
\emph{AI reallocated tasks, not headcount, in this project.} Delegation was highly uneven: implementation was almost fully AI-assisted, while architecture and prompt design was human-driven. Organizations may benefit from planning which activities to delegate rather than how many people to remove, with a \textit{small team} potentially absorbing more implementation work than headcount alone would suggest, provided judgment-heavy roles remain staffed. AI accelerated verbose more than conceptually complex tasks, a pattern story points may not capture. Finally, lightweight Architectural Description Records appeared to reduce context loss during long agent sessions, although we do not claim causation.
 
\noindent\textbf{For SE Education.}
Because this project originated in an AI-assisted software engineering course, it suggests, but does not validate, curricular directions. Classical engineering practices, such as writing specifications, documenting decisions, and critically reviewing AI outputs, appeared to reduce rework. Having students record costs, effort, and counterfactuals also exposed measurement biases, suggesting pedagogical value beyond the software itself.

\subsection{Limitations}
\label{sec:threats}
As emerging results from a single, ongoing case study, these findings are preliminary. The counterfactual estimates effort that was never actually performed, based on retrospective student judgment rather than a measured baseline — the same class of error that produced our initial 19.4× ratio, so even 9.9× may not be final. The team consisted of students, not the professionals used to price the counterfactual, and effort is self-reported. As a single case, the ratio is an early signal to be tested across more projects, not a benchmark. We ground verifiable terms (billing records, time logs) where possible, but the number and the methodology are work in progress.

\vspace{-2pt}
 
\section{Conclusion}
\label{sec:conclusion}
 
These early results suggest that pervasive AI assistance can shift the feasibility frontier for small development teams building non-trivial software: 35.2 hours of human effort and US\$69 of tooling produced a system that we estimate would otherwise have taken 329 professional hours. We suspect the measurement errors we corrected, inferring flat-rate costs from tokens and misapplying distant labor rates, are common, making our corrected methodology as valuable as the 9.9x ratio itself.

\section*{Artifact Availability}
\label{sec:availability}
A replication package\footnote{Available at: \url{https://doi.org/10.5281/zenodo.21843234}} includes the development log, counterfactual references, redacted billing records, prompts, and documented limitations (self-reported effort, estimated counterfactual hours, and apportioned subscription costs).
 
\section*{Generative AI Use Disclosure}
%In accordance with the venue's generative AI policy, 
The authors disclose that generative AI tools assisted in drafting text, structuring tables, and reformulating language. The authors reviewed, edited, and verified all generated text and tables. Additionally, the software system described in this paper was developed using pervasive AI assistance, as documented throughout the study.

\section*{Acknowledgements}
The authors thank the support of INES.IA (National Institute of Science and Technology for Software Engineering Based on and for Artificial Intelligence) www.ines.org.br, CNPq grant 408817/2024-0.

%% ------------------------------------------------------------
%% BIBLIOGRAPHY
%% ------------------------------------------------------------
\balance
\bibliographystyle{ACM-Reference-Format}
\bibliography{references}

\end{document}